\documentclass[aps,pre,reprint,superscriptaddress,longbibliography]{revtex4-1}
\usepackage{amsmath,amssymb,graphicx,subfigure,color,times,tabularx}

\begin{document}
\title{\bf {The Symmetry-Preserving Mean Field Condition for Electrostatic Correlations in Bulk}}
\author{Zhonghan Hu} \email{zhonghanhu@sdu.edu.cn}
\affiliation{Qingdao Institute for Theoretical and Computational Sciences (QiTCS), Shandong University, Qingdao, 266237, P. R. China}
\affiliation{Institute of Frontier and Interdisciplinary Science, Shandong University, Qingdao, 266237, P. R. China}
%\date{\today}
\begin{abstract}
Accurate simulations of a condensed system of ions or polar molecules are concerned with proper handlings of the involved electrostatics.
For such a Coulomb system at a charged planar interface, the Coulomb interaction averaged over the lateral directions with preserved symmetry serves as a necessary constraint in building any accurate handling that reconciles a simulated singlet charge density with the corresponding macroscopic charge/dielectric response.
At present, this symmetry-preserving mean-field (SPMF) condition represented in the reciprocal space, is conjectured to be necessary for a simulated bulk system to reproduce correctly the charge structure factor of the macroscopic bulk, as well.
In this work, we further examine analytically the asymptotic behavior of the charge structure factor at small wavenumbers for an arbitrary charge-charge interaction. In light of our theoretical predictions, simulations with a length of nearly $0.1$ micron are carried out to demonstrate that, typical efficient methods violating the
SPMF condition indeed fail to capture the exact charge correlations at small wavenumbers for both ionic and polar systems.
However, for both types of systems, these existing methods can be simply amended to match the SPMF condition and subsequently to probe precisely the electrostatic correlations at all length scales.
\end{abstract}
\maketitle

%%%%%%%%%%%%%%%%%%%%%%%%%%%%%%%%%%%%%%%%%%%%%%%%%%%%%%%%%%%%%%%%%%%%%%%%%%%%%%%%%%%%%%%%%%%%%%%%%%%%%%%%%%%%%%%%%%%%%%%%%%%%%%%%%%%%%%%%%%%%%%%%%%%%%%%%%%%%%%%%
There have been much advances regarding efficient and accurate treatments of the long-ranged electrostatics for molecular dynamics and Monte-Carlo simulations in recent
years\cite{Allen_Tildesley2017,Yi_Hu2015,Lowe_Sakata2018,Maggs2002,Fukuda2013,Wang_Fukuda2016,Wang2016,Liang_Xing2015,Girotto_Levin2017,Bakhshandeh_Levin2018,Zhang_E2018,Wang_E2018,Urano_Okazaki2020,Yuan_Luijten2021,Shi_Xu2021}.
These techniques differ from each other in various means:
for bulk\cite{Maggs2002,Fukuda2013,Wang_Fukuda2016,Wang2016,Zhang_E2018,Wang_E2018,Urano_Okazaki2020,Shi_Xu2021} versus for interfaces\cite{Liang_Xing2015,Girotto_Levin2017,Bakhshandeh_Levin2018,Yuan_Luijten2021}, 
efficient algorithms based on existing formulations\cite{Wang2016,Urano_Okazaki2020,Yuan_Luijten2021} versus newly designed functional forms of the simulated electrostatics\cite{Fukuda2013,Liang_Xing2015,Girotto_Levin2017,Bakhshandeh_Levin2018},
for classical non-polarizable systems\cite{Maggs2002,Fukuda2013,Wang_Fukuda2016,Wang2016,Urano_Okazaki2020,Shi_Xu2021} versus for systems with dielectric contrast\cite{Liang_Xing2015,Girotto_Levin2017,Bakhshandeh_Levin2018,Yuan_Luijten2021}, 
and methods well characterized by a simulation Hamiltonian in the framework of classical statistical mechanics\cite{Fukuda2013,Liang_Xing2015,Wang_Fukuda2016,Wang2016,Girotto_Levin2017,Bakhshandeh_Levin2018,Urano_Okazaki2020,Yuan_Luijten2021} versus others that might not be easily characterized in the same
way\cite{Maggs2002,Zhang_E2018,Wang_E2018,Shi_Xu2021}.
The development and implementation of these techniques have not only enhanced greatly our knowledge of complex Coulomb systems but also persistently called for a conceptual understanding of the intrinsic connections among the many techniques themselves from a transparent theoretical viewpoint.

The recent symmetry-preserving mean-field theory\cite{Hu2014spmf,Yi_Hu2017mf,Pan_Hu2017,Pan_Hu2019} might shed some light on the connections.
The core of this theory applied to the interfacial electrostatics is the so-called SPMF condition --- the lateral average of the pairwise charge-charge interaction, $\nu({\mathbf r})$ involved in an accurate simulation must equal that of the Coulomb interaction\cite{Pan_Hu2019}
\begin{equation} \left< \nu({\mathbf r}) \right>_{\rm sp}  \equiv \frac{1}{A}\iint dxdy\, \nu({\mathbf r}) =  \frac{-\lvert z\rvert}{2A\varepsilon_0}, \label{eq:spmf} \end{equation}
where $A=L_xL_y$ is the cross-sectional area of the simulation box and $\varepsilon_0$ is the vacuum permittivity.  $\left<\,\right>_{\rm sp}$ stands for the symmetry-preserving (sp) average over the lateral directions\cite{Hu2014spmf}. 
Whenever it operates on a two-dimensional (2D) Fourier series, integrations of trigonometric functions all vanish and only the first Fourier coefficient is left over\cite{Pan_Hu2019}.
$\nu({\mathbf r})$ can always be defined by rewriting the electrostatic energy in the simulation as\cite{Hu2014ib,Yi_Hu2017pairwise,Yuan_Luijten2021}
\begin{equation} {\cal U}_{\rm elec} = \sum_{i<j} q_i q_j \nu({\mathbf r}_{ij}),  \end{equation}
where $q_i$ is the charge of the $i$-th particle and ${\mathbf r}_{ij}={\mathbf r}_i - {\mathbf r}_j$ denotes the relative vector between the $i$- and $j$-th particles.
\begin{figure*}[!htb]\centerline{\includegraphics[width=15cm]{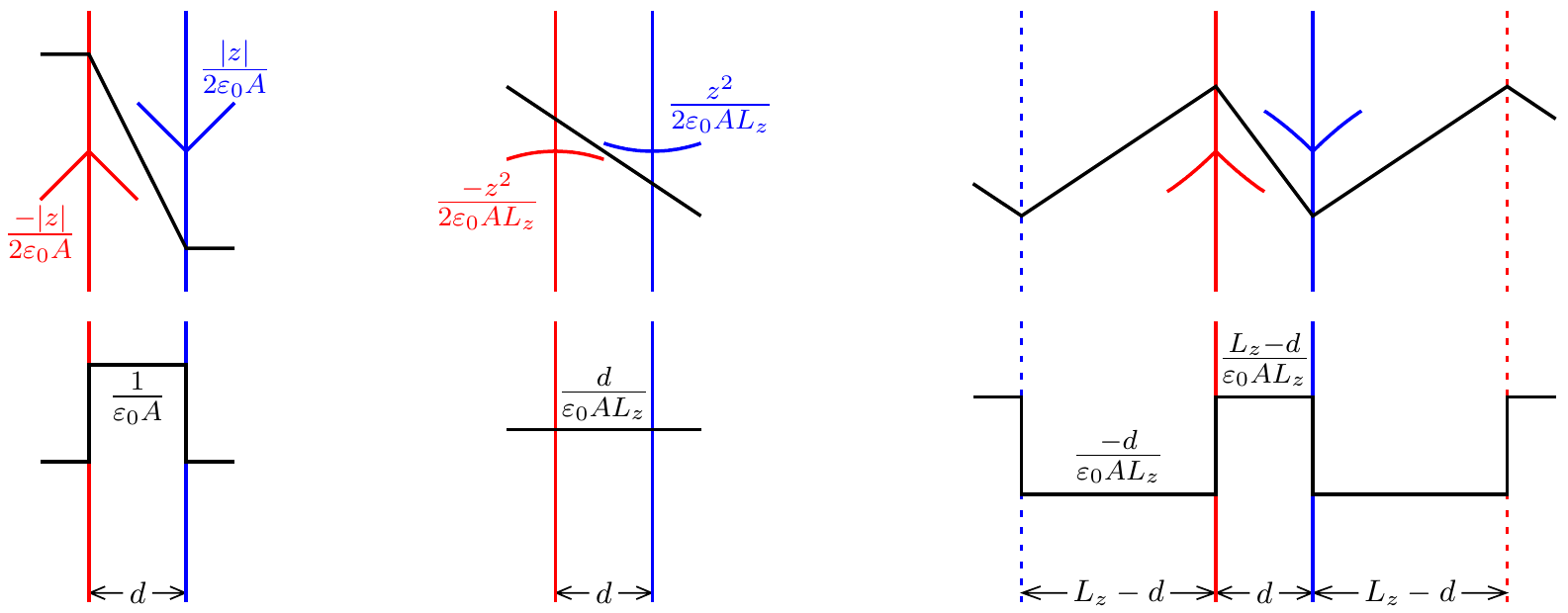} }
\caption{The electric potentials (top) and the corresponding electric fields (bottom) produced by a pair of unit charges (red solid lines for $q=+1$ and blue solid lines for $q=-1$) separated at $d$ interacting through the laterally averaged Coulomb interaction (left), the quadratic infinite boundary term (middle) and the periodic series (right) of
Eq.~\eqref{eq:spmfb}, respectively. Any potential (black) produced by the pair is the sum of the two individual potentials (colored).
The dash lines indicate the periodic images of the pair.}\label{fig:fig1}\end{figure*}

The SPMF condition is necessary for a finite simulation consist of a few thousands to billions of particles to reproduce the charge/dielectric response of the corresponding macroscopic Coulomb system ($\sim 10^{23}$ particles)\cite{Pan_Hu2019}.
To illustrate this statement precisely, let us consider an example of mobile ions confined between two macroscopic plate electrodes with fixed surface charges, $\pm\sigma$.
In this model system, the mobile ions driven by the electric field must form a nontrival charge distribution along the normal ($z$) direction in the interfacial region around each electrode.
At thermal equilibrium under normal conditions, the fixed surface charge of each electrode ($\pm\sigma$) should be screened completely by the compensating mobile surface charge ($\mp\sigma$) resulting from the integration of the equilibrium singlet charge density over the corresponding interfacial region, such that, mobile ions
in the bulk region feel no net electrostatic force on average.
In a numerical simulation employing $\nu({\mathbf r})$ for the charge-charge interaction under the usual 2D periodic boundary condition (PBC), the SPMF condition becomes necessary and sufficient for producing the exact compensating surface charge, $\mp\sigma$. 
On the other hand, the condition is necessary but insufficient for producing the detailed variation of the equilibrium singlet charge density along the $z$ direction.
However, as long as $\nu({\mathbf r})$ differs from $\nu^{\rm e2d}({\mathbf r})$ involved in the formally exact Ewald2D sum method\cite{Arnold_Holm2002b,Tuckerman_Martyna2002,Hu2014ib,Pan_Hu2014} by a slowly varying long-ranged component, the SPMF theory has suggested that interfacial structure, thermodynamics as well as
dynamics in the $z$ direction can often be obtained accurately from the approximated but much more efficient  method\cite{Hu2014spmf}.
In terms of $\nu({\mathbf r})$, all existing methods for interfacial electrostatics can be classified into three categories: the formally exact Ewald2D method and its efficient algorithms, approximated SPMF methods (e.g. \cite{Bengtsson1999,Zhang_Sprik2020} and more in ref.~\cite{Pan_Hu2019}), and other
methods which violate the SPMF condition but could be amended to work well for interfaces\cite{Pan_Hu2019}.

%In addition,  failure of the method in reproducing compensating %charges might be corrected analytically as done a couple of time before in refs.\cite{}. The SPMF of course neglect the lateral correlation resulting from the small scale fluctuations (e.g. Fig.1 inset of ref.\cite{})
While the guiding constraint, Eq.~\eqref{eq:spmf}, and the associated SPMF theory provide a transparent analysis of the interfacial electrostatics, the question arises as to whether or not the electrostatics in the bulk phase could be simply analyzed as well.
Since the equilibrium singlet charge density in bulk vanishes absolutely, the existence of any possible constraint must be related to the charge correlations in general. 
In practice, various methods with diverse $\nu({\mathbf r})$ have all proved successful for treating the local correlations of Coulomb systems\cite{Allen_Tildesley2017,Wolf1999,Wang_Fukuda2016,Chen_Weeks2004} and at first glance, it may appear very difficult to built any direct relation between $\nu({\mathbf r})$ and the
charge correlations.
In this work, as opposed to validating the usual correlations in the real space, we focus on the charge structure factor defined as the ensemble average in the reciprocal space\cite{Hansen_McDonald2006},
\begin{equation} S({\mathbf k}) = \frac{1}{V} \sum_{ij} q_i q_j \left< e^{i {\mathbf k}\cdot {\mathbf r}_{ij} }\right> , \label{eq:sk} \end{equation}
where $V=L_xL_yL_z$ is the volume of the cubic simulation cell. % and the relative vector ${\mathbf r}_{ij}={\mathbf r}_i -{\mathbf r}_j$. % stands for the relative vector between the $i$- and $j$-th particles.
$S({\mathbf k})$ at ${\mathbf k}=2\pi(k_x/L_x,k_y/L_y,k_z/L_z)$ with $k_x$, $k_y$ and $k_z$ all integers, probes the charge correlation at the discrete wave vector that is compatible with the reciprocal lattice of the finite simulation cell.
Correspondingly, the averaged Coulomb interaction represented in the reciprocal space is known to be a combination of a non-periodic quadratic term and a periodic Fourier series\cite{Pan_Hu2017,Pan_Hu2019},
\begin{equation} \frac{-\lvert z \rvert}{2\varepsilon_0 A} = \frac{-z^2}{2\varepsilon_0 V} + \frac{1}{V}\sum_{k_z\neq 0} \frac{e^{ikz}}{\varepsilon_0 k^2} , \label{eq:spmfb}  \end{equation}
where $ k = \lvert {\mathbf k} \rvert $ with ${\mathbf k}$ specified to be $(0,0,2\pi k_z/L_z)$ normal to the $xy$ plane. Eq.~\eqref{eq:spmfb} is valid up to a constant for any $z\in [-L_z, L_z]$.
The quadratic term, previously called the infinite boundary term, identifies with the excluded $k \to 0$ term of the periodic series, provided that the divergence caused by $1/k$ and $1/k^2$ are all removed\cite{Hu2014ib}. 

The electric field produced by this quadratic term for a pair of unit charges ($q = \pm 1$) is a constant proportional to the distance between the two charges in the prescribed direction (see Fig.~\ref{fig:fig1}).
Each quadratic term in an instantaneous configuration thus adds up to contribute a constant electric field proportional to the non-periodic (itinerant) total dipole moment.
Being the reality of a non-periodic quantity, the itinerant dipole moment possesses no translational invariance and becomes ill-defined upon the PBC transform of any non-zero charge.
As such, the quadratic infinite boundary term, as its name says, must be merely responsible for the effect of the macroscopic boundary, which is deemed irrelevant to any well-defined bulk property such as $S({\mathbf k})$ of Eq.~\eqref{eq:sk}. 

When focusing on the translationally invariant electrostatic correlations under the full PBC, one might exclude the non-periodic term and subsequently conjecture that the pairwise interaction accounting for an accurate $S({\mathbf k})$ with ${\mathbf k}=(0,0,2\pi k_z/L_z)$ satisfies necessarily, 
\begin{equation} \left< \nu({\mathbf r}) \right>_{\rm sp} = \frac{1}{V}\sum_{k_z\neq 0} \frac{e^{ikz}}{\varepsilon_0 k^2} ,\label{eq:spmfp} \end{equation}
%where $ k = \lvert {\mathbf k}\rvert $ with ${\mathbf k}$ takes $(0,0,2\pi k_z/L_z)$ as above.
which defines the SPMF condition for the electrostatics in the bulk phase. 
Similar constraints apply to symmetry-preserving averages in other directions for the purpose of determining precisely $S({\mathbf k})$ at the corresponding wave vectors.
Noting that $1/(\varepsilon_0 k^2)$ is the 3D Fourier transform of the Coulomb interaction and $\left<\,\right>_{\rm sp}$ removes all the 3D Fourier coefficients associated with $e^{i{\mathbf k}\cdot {\mathbf r}}$ subject to $k_x\neq 0$ or $k_y \neq 0$, this SPMF condition suggests essentially that,
any chosen lateral average of $\nu({\mathbf r})$ represented in the reciprocal space must equal that of the Coulomb interaction such that the finite simulation under the 3D PBC is able to reproduce correctly the macroscopic charge correlations in the corresponding normal direction of the reciprocal space.

It turns out to be informative to examine the existing methods for bulk electrostatics in terms of Eq.~\eqref{eq:spmfp}.
The pairwise interaction of the well known Ewald3D sum method with the tinfoil boundary condition (e3dtf) reads\cite{Hu2014ib,Yi_Hu2017pairwise,Pan_Hu2019}
\begin{equation}  \nu^{\rm e3dtf}({\mathbf r}) = \tau^{\rm 3D} + \frac{1}{V} \sum_{{\mathbf k}\neq{\mathbf 0}} \frac{e^{i{\mathbf k}\cdot{\mathbf r}} }{\varepsilon_0k^2}, \label{eq:pwnu}  \end{equation}
which indeed satisfies Eq.~\eqref{eq:spmfp} (up to the constant $\tau^{\rm 3D}$) in any direction. The e3dtf method is therefore considered to be formally exact for probing the electrostatic correlations in the reciprocal space. 
This shift of perspective resolves the long-time concern regarding the artificial anisotropy introduced by the e3dtf method\cite{Caillol1992,Allen_Tildesley2017,Yi_Hu2015}: charge correlations in the reciprocal space under the 3D PBC are as isotropic as they are on the 3D hypersurface of a 4D sphere\cite{Caillol1992}.
In another word, any anisotropy introduced by $ \nu^{\rm e3dtf}({\mathbf r})$ in the real space won't be problematic any more once the quantity of
interest is computed properly by the inverse Fourier transform of the corresponding ${\mathbf k}$-dependent ensemble average.
When the Ewald3D sum is associated with other boundary conditions that depend on the itinerant dipole moment\cite{DeLeeuw_Smith1980,Caillol1994,Hu2014spmf}, Caillol has argued
that the itinerant dipole moment is a decoupled collective variable that behaves as an independent harmonic oscillator for an electrolyte\cite{Caillol1994}, which well supports our exclusion of the redundant non-periodic infinite boundary term.

Distinct pairwise interactions involved in other useful methods in principle violate the SPMF condition. 
As an example, the recent zero-multipole (zm) method introduces deformed Coulomb interactions, $\nu({\mathbf r}|l,\alpha)$ parameterized by the order of the multipole moment $l$ and the damping factor $\alpha$\cite{Fukuda2013,Wang_Fukuda2016}. 
Its 3D Fourier transform, which is the Fourier coefficient of the corresponding pairwise interaction, differs (d) from $1/(\varepsilon_0 k^2)$ by
\begin{equation}  \hat{\nu}_{\rm d}({\mathbf k}|l,\alpha) = 1/(\varepsilon_0 k^2) -  \hat{\nu}({\mathbf k}|l,\alpha). \label{eq:nud} \end{equation}
For three typical sets of parameters: $l=0$ (zm0), $l=2$ and $\alpha=0$ (zm2), and $l=3$ and $\alpha=0$ (zm3)\cite{Fukuda2013,Wang_Fukuda2016}, the remaining Fourier coefficients read explicitly
\begin{equation} \hat{\nu}_{\rm d}({\mathbf k}|0,\alpha) = e^{-k^2/(4\alpha^2)} /(\varepsilon_0k^2), \end{equation}
\begin{equation} \hat{\nu}_{\rm d}({\mathbf k}|2,0) = \frac{15\left[ (3-k_c^2)\sin k_c -  3k_c\cos k_c\right]}{\varepsilon_0 k^2k_c^5}, \end{equation}
where $k_c=kr_c$ with $r_c$ the cutoff distance, 
and
\begin{equation} \hat{\nu}_{\rm d}({\mathbf k}|3,0) = \frac{105\left[ (15-6k_c^2)\sin k_c - (15-k_c^2)k_c \cos k_c \right]  }{\varepsilon_0 k^2k_c^7}, \label{eq:nud3}\end{equation}
respectively. $\hat{\nu}_{\rm d}({\mathbf k}|l,\alpha)$ is nonzero unless both $\alpha=0$ and $r_c\to \infty$.
To match the SPMF condition in a prescribed (e.g. $z$) direction for the general case of finite $\alpha$ and $r_c$, the amended zero-multipole (azm) method must include an extra electrostatic term 
\begin{equation}  {\cal U}_{\rm extra} = \sum_{i<j} q_i q_j \nu_{\rm sp}(z_{ij}|l,\alpha),  \end{equation}
where $\nu_{\rm sp}(z|l,\alpha)$ sums over the remaining Fourier coefficients in the direction with preserved symmetry as in Eq.~\eqref{eq:spmfp}
\begin{equation} \nu_{\rm sp}(z|l,\alpha) = \frac{1}{V}\sum_{k_z \neq 0} \hat{\nu}_{\rm d}({\mathbf k}|l,\alpha) e^{ikz}. \end{equation}
Besides, both electrostatic terms corresponding to $\nu_{\rm sp}(x|l,\alpha)$ and $\nu_{\rm sp}(y|l,\alpha)$ can be added to retain isotropy and subsequently match the SPMF condition in all three directions.

It is possible to examine analytically the influence of the different pairwise interactions on the charge structure factor, $S({\mathbf k})$ for both ionic and polar systems.
Because various methods have been successful in probing the short-ranged correlations\cite{Fukuda2013,Wang_Fukuda2016,Zhang_E2018,Wang_E2018,Shi_Xu2021}, which is related to $S({\mathbf k})$ at relatively large wavenumbers, deficiencies of the pairwise interaction violating the SPMF condition must be most
evident when one computes $S({\mathbf k})$ at small wavenumbers that reflect otherwise the long-ranged correlations.
For a conducting ionic fluid interacting through the Coulomb force, the seminal work of Stillinger and Lovett argued that $S(k)$ has the universal form at small wavenumbers\cite{Stillinger_Lovett1968}
\begin{equation} S({\mathbf k}) = 0 + k_b T \varepsilon_0 k^2 + {\cal O}(k^4) , \label{eq:sl}\end{equation}
where $k_b$ is the Boltzmann constant and $T$ is the temperature. This exact asymptotic form of course agrees with the well known Debye-Huckel limit\cite{Hansen_McDonald2006}
%\begin{equation} S({\mathbf k}) = k_b T \varepsilon_0/(\lambda^2 + \hat{\nu}({\mathbf k})\varepsilon_0 ),  \label{eq:dh} \end{equation}
\begin{equation} S({\mathbf k}) = \frac{k_b T \varepsilon_0}{\lambda^2 + \hat{\nu}({\mathbf k})\varepsilon_0 },  \label{eq:dh} \end{equation}
where $\hat{\nu}({\mathbf k})=1/(\varepsilon_0 k^2)$ and $\lambda$ is the Debye length defined through $\lambda^2 = k_bT\varepsilon_0/I$ with the ionic strength $I$ given by the sum of squares of charges divided by the volume.
Both forms in Eqs.~\eqref{eq:sl} and~\eqref{eq:dh} characterize the charge correlations in the conducting fluid --- any fixed charge is completely screened by the surrounding mobile charges as a consequence of the divergence of $\hat{\nu}({\mathbf k})$ at ${\mathbf k}\to 0$\cite{Hansen_McDonald2006}. 
This complete screening effect is irrespective of any details of the short-ranged non-electrostatic interaction.

In a simulation employing otherwise the deformed Coulomb interaction, the asymptotic behavior of $S({\mathbf k})$ can be readily obtained by replacing $\hat{\nu}({\mathbf k})$ in Eq.~\eqref{eq:dh} with $\hat{\nu}({\mathbf k}|l,\alpha)$, which however remains finite at ${\mathbf k}\to 0$ for finite $\alpha$ and $r_c$.
Therefore, any short-ranged pairwise interaction must lead to an incomplete screening which gives the divergence of the screening function, $S({\mathbf k})/k^2$ at ${\mathbf k}\to 0$: 
\begin{equation} \frac{S({\mathbf k})}{k^2} = \frac{k_b T \varepsilon_0}{\lambda^2 k^2 + \hat{\nu}({\mathbf k}|l,\alpha)\varepsilon_0k^2  } + {\cal O}(k^2).  \label{eq:sf1} \end{equation}
When the simulation box is large enough to evaluate $S({\mathbf k})$ of Eq.~\eqref{eq:sk} at a sufficient small $k$, $S({\mathbf k})/k^2$ for the simulated conducting fluid must approach the infinite.

\begin{figure}[!htb]\centerline{\includegraphics[width=7cm]{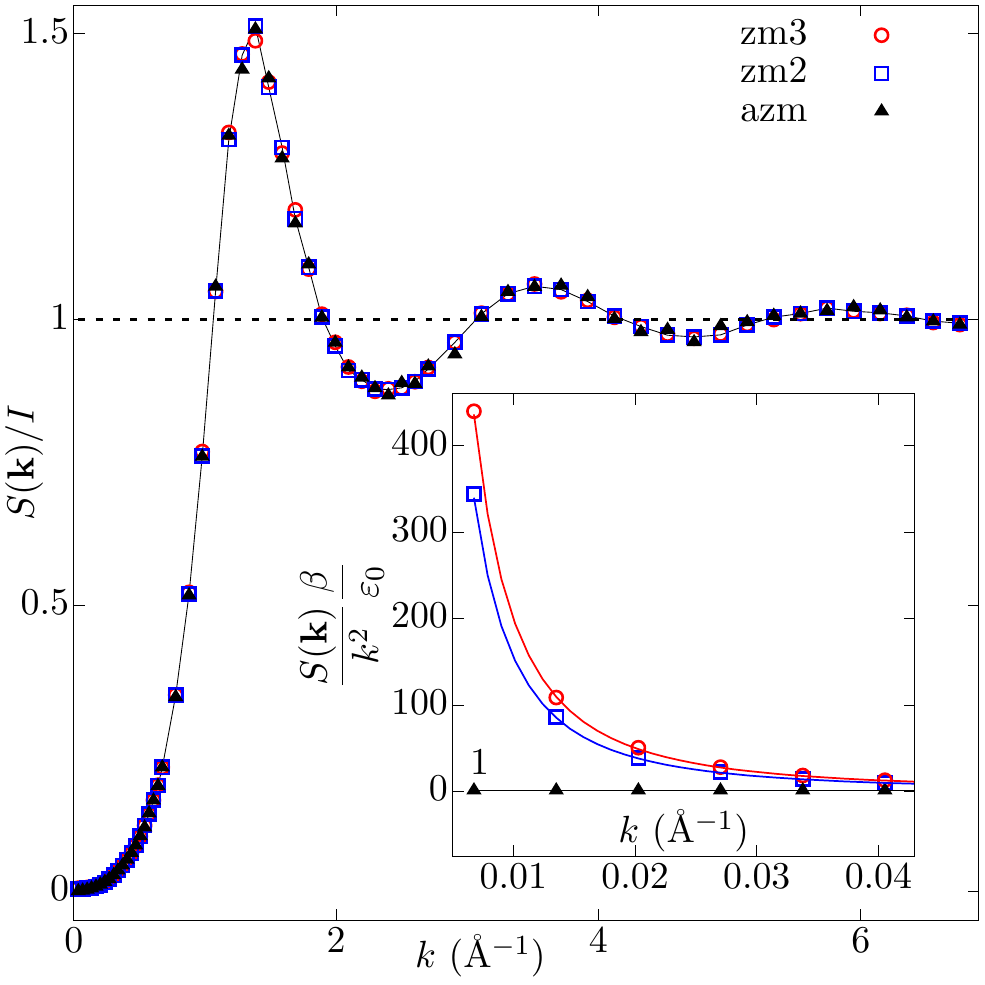} }
\caption{The charge structure factors and screening functions (inset) reduced by the ionic strength and $\varepsilon_0/\beta$ respectively for the ionic fluid simulated with the three zero-multiple methods. $\beta=1/(k_bT)$. The dash line and the solid lines (inset) indicate the analytic asymptotes at large and small wavenumbers
respectively. $\lambda=0.526\,\mbox{\AA}$}\label{fig:fig2}\end{figure}
\begin{figure}[!htb]\centerline{\includegraphics[width=7cm]{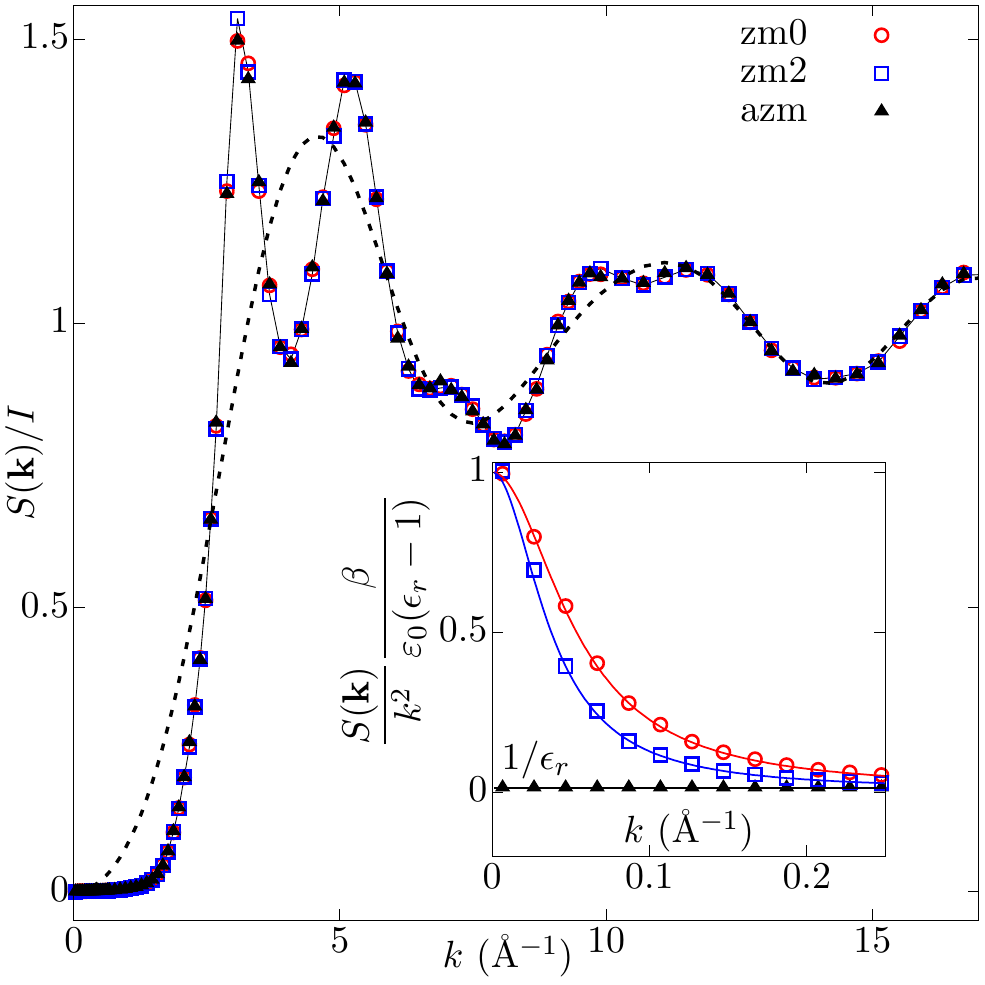} }
\caption{Same as Fig.~\ref{fig:fig2} but for the SPC/E model of water. The screening functions (inset) are additionally reduced by $\epsilon_r -1$ with $\epsilon_r\simeq 70$.}\label{fig:fig3}\end{figure}

Similar things happen to a molecular fluid with a dielectric constant (relative permittivity) of $\epsilon_r$ ($\epsilon_r \geqslant 1$). $\lambda^2k^2$ in Eq.~\eqref{eq:sf1} is substituted with $1/(\epsilon_r -1)$
\begin{equation} \frac{S({\mathbf k})}{k^2} = \frac{k_b T \varepsilon_0}{1/(\epsilon_r -1)+ \hat{\nu}({\mathbf k}|l,\alpha)\varepsilon_0k^2  } + {\cal O}(k^2),  \label{eq:sf2} \end{equation}
such that $S({\mathbf k})$ approaches the correct limit of the dielectric response\cite{Chandler1977,Rodgers_Weeks2009}
\begin{equation}  S({\mathbf k}) = k_bT \varepsilon_0 (\epsilon_r -1) k^2 /\epsilon_r + {\cal O}(k^4),  \end{equation}
when $\hat{\nu}({\mathbf k}|l,\alpha)$ approaches $1/(\varepsilon_0 k^2)$ at $\alpha\to0$ and $r_c\to\infty$.
The screening function evaluated in a simulation using the short-ranged $\nu({\mathbf r}|l,\alpha)$ now approaches the limit $k_bT\varepsilon_0(\epsilon_r-1)$. This limit, away from divergence for any finite $\epsilon_r$, is expected because there is no unscreened free charge any more in the molecular fluid.

The above analysis emphasizes that the SPMF condition imposed on the pairwise interaction is a direct consequence of the screening effect of the Coulomb force. Violation of the SPMF condition leads to improper behaviors of charge correlations characterized by $S({\mathbf k})/k^2$ at small wavenumbers.
In light of these theoretical predications, we carry out simulations of the model ionic fluid\cite{Hu_Weeks2010lmf} at $T=5000\,K$ and the SPC/E water\cite{Berendsen_Straatsma1987} at $T=298.15\,K$ with one length of the simulation box $930 \,\mbox{\AA}$ and $939.6 \,\mbox{\AA}$, respectively. 
In the zm2/zm3 methods, $r_c=30 \,\mbox{\AA}$ and $r_c=12\, \mbox{\AA}$ for the ionic fluid and the water respectively. In the zm0 method, $\alpha =1/4.5 \,\mbox{\AA}^{-1}$ as in other simulations of water\cite{Hu_Weeks2010lmf,Remsing_Weeks2016,Gao_Weeks2020}.
The ionic strengths $I=0.00859  \,e^2\mbox{\AA}^{-3}$ and  $I=0.0359 \, e^2\mbox{\AA}^{-3}$ for the ionic fluid and the water respectively. The Debye length of the ionic fluid is $\lambda = 0.526\, \mbox{\AA}$. Complete details of the simulations are available to the public\cite{notedetails}.

Figs.~\ref{fig:fig2} and~\ref{fig:fig3} display the computed $S({\mathbf k})$ and $S({\mathbf k})/k^2$ at discrete wavenumbers for both systems.
At extremely large $k$, $S({\mathbf k})$ approaches the corresponding ionic strength indicating the self-correlation.  In addition, the damped oscillation of $S({\mathbf k})$ for the SPC/E water approaches the intra-molecular correlation
\begin{equation} \frac{S({\mathbf k})}{I}\sim 1 + \frac{4q_{\rm O}q_{\rm H}}{q_{\rm sum}} \frac{\sin(kd_{\rm OH})}{kd_{\rm OH}} +\frac{2q_{\rm H}^2}{q_{\rm sum}}  \frac{\sin(kd_{\rm HH})}{kd_{\rm HH}} ,\end{equation} 
%\begin{equation} \frac{S({\mathbf k})}{I}\sim 1 + \frac{4q_{\rm O}q_{\rm H}}{q_{\rm O}^2 + 2 q_{\rm H}^2} \frac{\sin(kd_{\rm OH})}{kd_{\rm OH}} +\frac{4q_{\rm O}q_{\rm H}}{q_{\rm O}^2 + 2 q_{\rm H}^2}  \frac{\sin(kd_{\rm HH})}{kd_{\rm HH}} ,\end{equation} 
%\begin{equation} \frac{S({\mathbf k})}{I}\sim 1 + \frac{ 4q_{\rm O}q_{\rm H} d_{\rm HH} \sin(kd_{\rm OH}) + 2 q_{\rm H}^2 d_{\rm OH} \sin(kd_{\rm HH}) }{ (q_{\rm O}^2 + 2 q_{\rm H}^2) d_{\rm OH} d_{\rm HH} k } ,\end{equation}
where $q_{\rm O}$, $q_{\rm H}$, $d_{\rm OH}$ and $d_{\rm HH}$ are the parameters of charges and bond lengths in the SPC/E model\cite{Berendsen_Straatsma1987,notedetails}.
%: $q_{\rm O} = -0.8476\,e$, $q_{\rm H} = 0.4238 \, e$, $d_{\rm OH} = 1\,\mbox{\AA}$, and $d_{\rm HH} = 1.633 \,\mbox{\AA}$. 
$q_{\rm sum} = q_{\rm O}^2 + 2q_{\rm H}^2$, is the sum of the squares of the charges in one molecule.
$S({\mathbf k})$ from all methods are indistinguishable at the wavenumbers corresponding to distances ($2\pi/k$) smaller than the local correlation lengths. However, distinct features between the SPMF condition satisfied and violated methods are found at small $k$ for both the ionic fluid and the water. 
These characteristics are in excellent agreement with the analytical predictions from Eqs.~\eqref{eq:sf1} and~\eqref{eq:sf2} with $\hat{\nu}({\mathbf k}|l,\alpha)$ determined by Eqs.~\eqref{eq:nud} to~\eqref{eq:nud3} and the specified parameters of the systems.

In conclusion, the SPMF condition provides a simple concept to connect the electrostatic correlations among Coulomb and simulated Coulomb systems. 
Exceptions are simulations employing techniques that are not well represented by a pairwise interaction. These useful methods include the local maxwell approach\cite{Maggs2002}, the random batch method\cite{Shi_Xu2021,Liang_Xu2021} and the fast growing machine-learning techniques\cite{Zhang_E2018,Wang_E2018}. 
While analytic understandings of the mentioned methods have yet to be achieved, the present work suggests that it is convenient to validate them by calculating correlations in the reciprocal space such as $S({\mathbf k})$, the usual structure factor and intermediate scattering functions\cite{Hansen_McDonald2006}.
We note that numerical evidence does show that the local maxwell approach behaves correctly at small wavenumbers\cite{Maggs2002}.
Further investigations at all wavelengths for both time-averaged and time-displaced correlations are expected.
We finally hope that the SPMF condition for interfacial and bulk electrostatics helps to provide a simple starting point when developing methods to simulate a complex condensed phase.

This work was supported by the NSFC (Grant No. 21873037)
%%%%%%%%%%%%%%%%%%%%%%%%%%%%%%%%%%%%%%%%%%%%%%%%%%%%%%%%%%%%%%%%%%%%%%%%%%%%%%%%%%%%%%%%%%%%%%%%%%%%%%%%%%%%%%%%%%%%%%%%%%%%%%%%%%%%%%%%%%%%%%%%%%%%%%%%%%%%%%%%

%\bibliography{refs1990b,refs1990,refs2000,refs2010,refs2020,refsnote}

\begin{thebibliography}{41}%
\makeatletter
\providecommand \@ifxundefined [1]{%
 \@ifx{#1\undefined}
}%
\providecommand \@ifnum [1]{%
 \ifnum #1\expandafter \@firstoftwo
 \else \expandafter \@secondoftwo
 \fi
}%
\providecommand \@ifx [1]{%
 \ifx #1\expandafter \@firstoftwo
 \else \expandafter \@secondoftwo
 \fi
}%
\providecommand \natexlab [1]{#1}%
\providecommand \enquote  [1]{``#1''}%
\providecommand \bibnamefont  [1]{#1}%
\providecommand \bibfnamefont [1]{#1}%
\providecommand \citenamefont [1]{#1}%
\providecommand \href@noop [0]{\@secondoftwo}%
\providecommand \href [0]{\begingroup \@sanitize@url \@href}%
\providecommand \@href[1]{\@@startlink{#1}\@@href}%
\providecommand \@@href[1]{\endgroup#1\@@endlink}%
\providecommand \@sanitize@url [0]{\catcode `\\12\catcode `\$12\catcode
  `\&12\catcode `\#12\catcode `\^12\catcode `\_12\catcode `\%12\relax}%
\providecommand \@@startlink[1]{}%
\providecommand \@@endlink[0]{}%
\providecommand \url  [0]{\begingroup\@sanitize@url \@url }%
\providecommand \@url [1]{\endgroup\@href {#1}{\urlprefix }}%
\providecommand \urlprefix  [0]{URL }%
\providecommand \Eprint [0]{\href }%
\providecommand \doibase [0]{http://dx.doi.org/}%
\providecommand \selectlanguage [0]{\@gobble}%
\providecommand \bibinfo  [0]{\@secondoftwo}%
\providecommand \bibfield  [0]{\@secondoftwo}%
\providecommand \translation [1]{[#1]}%
\providecommand \BibitemOpen [0]{}%
\providecommand \bibitemStop [0]{}%
\providecommand \bibitemNoStop [0]{.\EOS\space}%
\providecommand \EOS [0]{\spacefactor3000\relax}%
\providecommand \BibitemShut  [1]{\csname bibitem#1\endcsname}%
\let\auto@bib@innerbib\@empty
%</preamble>
\bibitem [{\citenamefont {Allen}\ and\ \citenamefont
  {Tildesley}(2017)}]{Allen_Tildesley2017}%
  \BibitemOpen
  \bibfield  {author} {\bibinfo {author} {\bibfnamefont {M.~P.}\ \bibnamefont
  {Allen}}\ and\ \bibinfo {author} {\bibfnamefont {D.~J.}\ \bibnamefont
  {Tildesley}},\ }\href@noop {} {\emph {\bibinfo {title} {Computer Simulation
  of Liquids}}},\ \bibinfo {edition} {2nd}\ ed.\ (\bibinfo  {publisher} {Oxford
  University Press, Oxford, UK},\ \bibinfo {year} {2017})\BibitemShut {NoStop}%
\bibitem [{\citenamefont {Yi}\ \emph {et~al.}(2015)\citenamefont {Yi},
  \citenamefont {Pan},\ and\ \citenamefont {Hu}}]{Yi_Hu2015}%
  \BibitemOpen
  \bibfield  {author} {\bibinfo {author} {\bibfnamefont {Shasha}\ \bibnamefont
  {Yi}}, \bibinfo {author} {\bibfnamefont {Cong}\ \bibnamefont {Pan}}, \ and\
  \bibinfo {author} {\bibfnamefont {Zhonghan}\ \bibnamefont {Hu}},\ }\bibfield
  {title} {\enquote {\bibinfo {title} {Accurate treatments of electrostatics
  for computer simulations of biological systems: A brief survey of
  developments and existing problems},}\ }\href@noop {} {\bibfield  {journal}
  {\bibinfo  {journal} {Chin. Phys. B}\ }\textbf {\bibinfo {volume} {24}},\
  \bibinfo {pages} {120201} (\bibinfo {year} {2015})}\BibitemShut {NoStop}%
\bibitem [{\citenamefont {Lowe}\ \emph {et~al.}(2018)\citenamefont {Lowe},
  \citenamefont {Skylaris}, \citenamefont {Green}, \citenamefont {Shibuta},\
  and\ \citenamefont {Sakata}}]{Lowe_Sakata2018}%
  \BibitemOpen
  \bibfield  {author} {\bibinfo {author} {\bibfnamefont {Benjamin~M.}\
  \bibnamefont {Lowe}}, \bibinfo {author} {\bibfnamefont {Chris-Kriton}\
  \bibnamefont {Skylaris}}, \bibinfo {author} {\bibfnamefont {Nicolas~G.}\
  \bibnamefont {Green}}, \bibinfo {author} {\bibfnamefont {Yasushi}\
  \bibnamefont {Shibuta}}, \ and\ \bibinfo {author} {\bibfnamefont {Toshiya}\
  \bibnamefont {Sakata}},\ }\bibfield  {title} {\enquote {\bibinfo {title}
  {Calculation of surface potentials at the silica{\textendash}water interface
  using molecular dynamics: Challenges and opportunities},}\ }\href@noop {}
  {\bibfield  {journal} {\bibinfo  {journal} {Japanese J. App. Phys.}\ }\textbf
  {\bibinfo {volume} {57}},\ \bibinfo {pages} {04FM02} (\bibinfo {year}
  {2018})}\BibitemShut {NoStop}%
\bibitem [{\citenamefont {Maggs}(2002)}]{Maggs2002}%
  \BibitemOpen
  \bibfield  {author} {\bibinfo {author} {\bibfnamefont {A.~C.}\ \bibnamefont
  {Maggs}},\ }\bibfield  {title} {\enquote {\bibinfo {title} {Dynamics of a
  local algorithm for simulating coulomb interactions},}\ }\href@noop {}
  {\bibfield  {journal} {\bibinfo  {journal} {J. Chem. Phys.}\ }\textbf
  {\bibinfo {volume} {117}},\ \bibinfo {pages} {1975--1981} (\bibinfo {year}
  {2002})}\BibitemShut {NoStop}%
\bibitem [{\citenamefont {Fukuda}(2013)}]{Fukuda2013}%
  \BibitemOpen
  \bibfield  {author} {\bibinfo {author} {\bibfnamefont {Ikuo}\ \bibnamefont
  {Fukuda}},\ }\bibfield  {title} {\enquote {\bibinfo {title} {Zero-multipole
  summation method for efficiently estimating electrostatic interactions in
  molecular system},}\ }\href@noop {} {\bibfield  {journal} {\bibinfo
  {journal} {J. Chem. Phys.}\ }\textbf {\bibinfo {volume} {139}},\ \bibinfo
  {pages} {174107} (\bibinfo {year} {2013})}\BibitemShut {NoStop}%
\bibitem [{\citenamefont {Wang}\ \emph
  {et~al.}(2016{\natexlab{a}})\citenamefont {Wang}, \citenamefont {Nakamura},\
  and\ \citenamefont {Fukuda}}]{Wang_Fukuda2016}%
  \BibitemOpen
  \bibfield  {author} {\bibinfo {author} {\bibfnamefont {Han}\ \bibnamefont
  {Wang}}, \bibinfo {author} {\bibfnamefont {Haruki}\ \bibnamefont {Nakamura}},
  \ and\ \bibinfo {author} {\bibfnamefont {Ikuo}\ \bibnamefont {Fukuda}},\
  }\bibfield  {title} {\enquote {\bibinfo {title} {A critical appraisal of the
  zero-multipole method: Structural, thermodynamic, dielectric, and dynamical
  properties of a water system},}\ }\href@noop {} {\bibfield  {journal}
  {\bibinfo  {journal} {J. Chem. Phys.}\ }\textbf {\bibinfo {volume} {144}},\
  \bibinfo {pages} {114503} (\bibinfo {year} {2016}{\natexlab{a}})}\BibitemShut
  {NoStop}%
\bibitem [{\citenamefont {Wang}\ \emph
  {et~al.}(2016{\natexlab{b}})\citenamefont {Wang}, \citenamefont {Gao},\ and\
  \citenamefont {Fang}}]{Wang2016}%
  \BibitemOpen
  \bibfield  {author} {\bibinfo {author} {\bibfnamefont {Han}\ \bibnamefont
  {Wang}}, \bibinfo {author} {\bibfnamefont {Xingyu}\ \bibnamefont {Gao}}, \
  and\ \bibinfo {author} {\bibfnamefont {Jun}\ \bibnamefont {Fang}},\
  }\bibfield  {title} {\enquote {\bibinfo {title} {Multiple staggered mesh
  ewald: Boosting the accuracy of the smooth particle mesh ewald method},}\
  }\href@noop {} {\bibfield  {journal} {\bibinfo  {journal} {J Chem. Theory
  Comput.}\ }\textbf {\bibinfo {volume} {12}},\ \bibinfo {pages} {5596--5608}
  (\bibinfo {year} {2016}{\natexlab{b}})}\BibitemShut {NoStop}%
\bibitem [{\citenamefont {Liang}\ \emph {et~al.}(2015)\citenamefont {Liang},
  \citenamefont {Xu},\ and\ \citenamefont {Xing}}]{Liang_Xing2015}%
  \BibitemOpen
  \bibfield  {author} {\bibinfo {author} {\bibfnamefont {Yihao}\ \bibnamefont
  {Liang}}, \bibinfo {author} {\bibfnamefont {Zhenli}\ \bibnamefont {Xu}}, \
  and\ \bibinfo {author} {\bibfnamefont {Xiangjun}\ \bibnamefont {Xing}},\
  }\bibfield  {title} {\enquote {\bibinfo {title} {A multi-scale monte carlo
  method for electrolytes},}\ }\href@noop {} {\bibfield  {journal} {\bibinfo
  {journal} {New J. Phys.}\ }\textbf {\bibinfo {volume} {17}},\ \bibinfo
  {pages} {083062} (\bibinfo {year} {2015})}\BibitemShut {NoStop}%
\bibitem [{\citenamefont {Girotto}\ \emph {et~al.}(2017)\citenamefont
  {Girotto}, \citenamefont {dos Santos},\ and\ \citenamefont
  {Levin}}]{Girotto_Levin2017}%
  \BibitemOpen
  \bibfield  {author} {\bibinfo {author} {\bibfnamefont {Matheus}\ \bibnamefont
  {Girotto}}, \bibinfo {author} {\bibfnamefont {Alexandre~P.}\ \bibnamefont
  {dos Santos}}, \ and\ \bibinfo {author} {\bibfnamefont {Yan}\ \bibnamefont
  {Levin}},\ }\bibfield  {title} {\enquote {\bibinfo {title} {Simulations of
  ionic liquids confined by metal electrodes using periodic green functions},}\
  }\href@noop {} {\bibfield  {journal} {\bibinfo  {journal} {J. Chem. Phys.}\
  }\textbf {\bibinfo {volume} {147}},\ \bibinfo {pages} {074109} (\bibinfo
  {year} {2017})}\BibitemShut {NoStop}%
\bibitem [{\citenamefont {Bakhshandeh}\ \emph {et~al.}(2018)\citenamefont
  {Bakhshandeh}, \citenamefont {dos Santos},\ and\ \citenamefont
  {Levin}}]{Bakhshandeh_Levin2018}%
  \BibitemOpen
  \bibfield  {author} {\bibinfo {author} {\bibfnamefont {Amin}\ \bibnamefont
  {Bakhshandeh}}, \bibinfo {author} {\bibfnamefont {Alexandre~P.}\ \bibnamefont
  {dos Santos}}, \ and\ \bibinfo {author} {\bibfnamefont {Yan}\ \bibnamefont
  {Levin}},\ }\bibfield  {title} {\enquote {\bibinfo {title} {Efficient
  simulation method for nano-patterned charged surfaces in an electrolyte
  solution},}\ }\href@noop {} {\bibfield  {journal} {\bibinfo  {journal} {Soft
  Matter}\ }\textbf {\bibinfo {volume} {14}},\ \bibinfo {pages} {4081--4086}
  (\bibinfo {year} {2018})}\BibitemShut {NoStop}%
\bibitem [{\citenamefont {Zhang}\ \emph {et~al.}(2018)\citenamefont {Zhang},
  \citenamefont {Han}, \citenamefont {Wang}, \citenamefont {Car},\ and\
  \citenamefont {E}}]{Zhang_E2018}%
  \BibitemOpen
  \bibfield  {author} {\bibinfo {author} {\bibfnamefont {Linfeng}\ \bibnamefont
  {Zhang}}, \bibinfo {author} {\bibfnamefont {Jiequn}\ \bibnamefont {Han}},
  \bibinfo {author} {\bibfnamefont {Han}\ \bibnamefont {Wang}}, \bibinfo
  {author} {\bibfnamefont {Roberto}\ \bibnamefont {Car}}, \ and\ \bibinfo
  {author} {\bibfnamefont {Weinan}\ \bibnamefont {E}},\ }\bibfield  {title}
  {\enquote {\bibinfo {title} {Deep potential molecular dynamics: A scalable
  model with the accuracy of quantum mechanics},}\ }\href@noop {} {\bibfield
  {journal} {\bibinfo  {journal} {Phys. Rev. Lett.}\ }\textbf {\bibinfo
  {volume} {120}},\ \bibinfo {pages} {143001} (\bibinfo {year}
  {2018})}\BibitemShut {NoStop}%
\bibitem [{\citenamefont {Wang}\ \emph {et~al.}(2018)\citenamefont {Wang},
  \citenamefont {Zhang}, \citenamefont {Han},\ and\ \citenamefont
  {E}}]{Wang_E2018}%
  \BibitemOpen
  \bibfield  {author} {\bibinfo {author} {\bibfnamefont {Han}\ \bibnamefont
  {Wang}}, \bibinfo {author} {\bibfnamefont {Linfeng}\ \bibnamefont {Zhang}},
  \bibinfo {author} {\bibfnamefont {Jiequn}\ \bibnamefont {Han}}, \ and\
  \bibinfo {author} {\bibfnamefont {Weinan}\ \bibnamefont {E}},\ }\bibfield
  {title} {\enquote {\bibinfo {title} {Deepmd-kit: A deep learning package for
  many-body potential energy representation and molecular dynamics},}\
  }\href@noop {} {\bibfield  {journal} {\bibinfo  {journal} {Computer Phys.
  Commun.}\ }\textbf {\bibinfo {volume} {228}},\ \bibinfo {pages} {178--184}
  (\bibinfo {year} {2018})}\BibitemShut {NoStop}%
\bibitem [{\citenamefont {Urano}\ \emph {et~al.}(2020)\citenamefont {Urano},
  \citenamefont {Shinoda}, \citenamefont {Yoshii},\ and\ \citenamefont
  {Okazaki}}]{Urano_Okazaki2020}%
  \BibitemOpen
  \bibfield  {author} {\bibinfo {author} {\bibfnamefont {Ryo}\ \bibnamefont
  {Urano}}, \bibinfo {author} {\bibfnamefont {Wataru}\ \bibnamefont {Shinoda}},
  \bibinfo {author} {\bibfnamefont {Noriyuki}\ \bibnamefont {Yoshii}}, \ and\
  \bibinfo {author} {\bibfnamefont {Susumu}\ \bibnamefont {Okazaki}},\
  }\bibfield  {title} {\enquote {\bibinfo {title} {Exact long-range coulombic
  energy calculation for net charged systems neutralized by uniformly
  distributed background charge using fast multipole method and its application
  to efficient free energy calculation},}\ }\href@noop {} {\bibfield  {journal}
  {\bibinfo  {journal} {J. Chem. Phys.}\ }\textbf {\bibinfo {volume} {152}},\
  \bibinfo {pages} {244115} (\bibinfo {year} {2020})}\BibitemShut {NoStop}%
\bibitem [{\citenamefont {Yuan}\ \emph {et~al.}(2021)\citenamefont {Yuan},
  \citenamefont {Antila},\ and\ \citenamefont {Luijten}}]{Yuan_Luijten2021}%
  \BibitemOpen
  \bibfield  {author} {\bibinfo {author} {\bibfnamefont {Jiaxing}\ \bibnamefont
  {Yuan}}, \bibinfo {author} {\bibfnamefont {Hanne~S.}\ \bibnamefont {Antila}},
  \ and\ \bibinfo {author} {\bibfnamefont {Erik}\ \bibnamefont {Luijten}},\
  }\bibfield  {title} {\enquote {\bibinfo {title} {Particle–particle
  particle–mesh algorithm for electrolytes between charged dielectric
  interfaces},}\ }\href@noop {} {\bibfield  {journal} {\bibinfo  {journal} {J.
  Chem. Phys.}\ }\textbf {\bibinfo {volume} {154}},\ \bibinfo {pages} {094115}
  (\bibinfo {year} {2021})}\BibitemShut {NoStop}%
\bibitem [{\citenamefont {Jin}\ \emph {et~al.}(2021)\citenamefont {Jin},
  \citenamefont {Li}, \citenamefont {Xu},\ and\ \citenamefont
  {Zhao}}]{Shi_Xu2021}%
  \BibitemOpen
  \bibfield  {author} {\bibinfo {author} {\bibfnamefont {Shi}\ \bibnamefont
  {Jin}}, \bibinfo {author} {\bibfnamefont {Lei}\ \bibnamefont {Li}}, \bibinfo
  {author} {\bibfnamefont {Zhenli}\ \bibnamefont {Xu}}, \ and\ \bibinfo
  {author} {\bibfnamefont {Yue}\ \bibnamefont {Zhao}},\ }\bibfield  {title}
  {\enquote {\bibinfo {title} {A random batch ewald method for particle systems
  with coulomb interactions},}\ }\href@noop {} {\bibfield  {journal} {\bibinfo
  {journal} {SIAM J. Sci. Comput.}\ }\textbf {\bibinfo {volume} {43}},\
  \bibinfo {pages} {B937--B960} (\bibinfo {year} {2021})}\BibitemShut {NoStop}%
\bibitem [{\citenamefont {Hu}(2014{\natexlab{a}})}]{Hu2014spmf}%
  \BibitemOpen
  \bibfield  {author} {\bibinfo {author} {\bibfnamefont {Zhonghan}\
  \bibnamefont {Hu}},\ }\bibfield  {title} {\enquote {\bibinfo {title}
  {Symmetry-preserving mean field theory for electrostatics at interfaces},}\
  }\href@noop {} {\bibfield  {journal} {\bibinfo  {journal} {Chem. Commun.}\
  }\textbf {\bibinfo {volume} {50}},\ \bibinfo {pages} {14397--14400} (\bibinfo
  {year} {2014}{\natexlab{a}})}\BibitemShut {NoStop}%
\bibitem [{\citenamefont {Yi}\ \emph {et~al.}(2017{\natexlab{a}})\citenamefont
  {Yi}, \citenamefont {Pan}, \citenamefont {Hu},\ and\ \citenamefont
  {Hu}}]{Yi_Hu2017mf}%
  \BibitemOpen
  \bibfield  {author} {\bibinfo {author} {\bibfnamefont {Shasha}\ \bibnamefont
  {Yi}}, \bibinfo {author} {\bibfnamefont {Cong}\ \bibnamefont {Pan}}, \bibinfo
  {author} {\bibfnamefont {Liming}\ \bibnamefont {Hu}}, \ and\ \bibinfo
  {author} {\bibfnamefont {Zhonghan}\ \bibnamefont {Hu}},\ }\bibfield  {title}
  {\enquote {\bibinfo {title} {On the connections and differences among three
  mean-field approximations: a stringent test},}\ }\href@noop {} {\bibfield
  {journal} {\bibinfo  {journal} {Phys. Chem. Chem. Phys.}\ }\textbf {\bibinfo
  {volume} {19}},\ \bibinfo {pages} {18514--18518} (\bibinfo {year}
  {2017}{\natexlab{a}})}\BibitemShut {NoStop}%
\bibitem [{\citenamefont {Pan}\ \emph {et~al.}(2017)\citenamefont {Pan},
  \citenamefont {Yi},\ and\ \citenamefont {Hu}}]{Pan_Hu2017}%
  \BibitemOpen
  \bibfield  {author} {\bibinfo {author} {\bibfnamefont {Cong}\ \bibnamefont
  {Pan}}, \bibinfo {author} {\bibfnamefont {Shasha}\ \bibnamefont {Yi}}, \ and\
  \bibinfo {author} {\bibfnamefont {Zhonghan}\ \bibnamefont {Hu}},\ }\bibfield
  {title} {\enquote {\bibinfo {title} {The effect of electrostatic boundaries
  in molecular simulations: symmetry matters},}\ }\href@noop {} {\bibfield
  {journal} {\bibinfo  {journal} {Phys. Chem. Chem. Phys.}\ }\textbf {\bibinfo
  {volume} {19}},\ \bibinfo {pages} {4861} (\bibinfo {year}
  {2017})}\BibitemShut {NoStop}%
\bibitem [{\citenamefont {Pan}\ \emph {et~al.}(2019)\citenamefont {Pan},
  \citenamefont {Yi},\ and\ \citenamefont {Hu}}]{Pan_Hu2019}%
  \BibitemOpen
  \bibfield  {author} {\bibinfo {author} {\bibfnamefont {Cong}\ \bibnamefont
  {Pan}}, \bibinfo {author} {\bibfnamefont {Shasha}\ \bibnamefont {Yi}}, \ and\
  \bibinfo {author} {\bibfnamefont {Zhonghan}\ \bibnamefont {Hu}},\ }\bibfield
  {title} {\enquote {\bibinfo {title} {Analytic theory of finite-size effects
  in supercell modelling of charged interfaces},}\ }\href@noop {} {\bibfield
  {journal} {\bibinfo  {journal} {Phys. Chem. Chem. Phys.}\ }\textbf {\bibinfo
  {volume} {21}},\ \bibinfo {pages} {14858} (\bibinfo {year}
  {2019})}\BibitemShut {NoStop}%
\bibitem [{\citenamefont {Hu}(2014{\natexlab{b}})}]{Hu2014ib}%
  \BibitemOpen
  \bibfield  {author} {\bibinfo {author} {\bibfnamefont {Zhonghan}\
  \bibnamefont {Hu}},\ }\bibfield  {title} {\enquote {\bibinfo {title}
  {Infinite boundary terms of ewald sums and pairwise interactions for
  electrostatics in bulk and at interfaces},}\ }\href@noop {} {\bibfield
  {journal} {\bibinfo  {journal} {J. Chem. Theory Comput.}\ }\textbf {\bibinfo
  {volume} {10}},\ \bibinfo {pages} {5254--5264} (\bibinfo {year}
  {2014}{\natexlab{b}})}\BibitemShut {NoStop}%
\bibitem [{\citenamefont {Yi}\ \emph {et~al.}(2017{\natexlab{b}})\citenamefont
  {Yi}, \citenamefont {Pan},\ and\ \citenamefont {Hu}}]{Yi_Hu2017pairwise}%
  \BibitemOpen
  \bibfield  {author} {\bibinfo {author} {\bibfnamefont {Shasha}\ \bibnamefont
  {Yi}}, \bibinfo {author} {\bibfnamefont {Cong}\ \bibnamefont {Pan}}, \ and\
  \bibinfo {author} {\bibfnamefont {Zhonghan}\ \bibnamefont {Hu}},\ }\bibfield
  {title} {\enquote {\bibinfo {title} {Note: A pairwise form of the ewald sum
  for non-neutral systems},}\ }\href@noop {} {\bibfield  {journal} {\bibinfo
  {journal} {J. Chem. Phys.}\ }\textbf {\bibinfo {volume} {147}},\ \bibinfo
  {pages} {126101} (\bibinfo {year} {2017}{\natexlab{b}})}\BibitemShut
  {NoStop}%
\bibitem [{\citenamefont {Arnold}\ \emph {et~al.}(2002)\citenamefont {Arnold},
  \citenamefont {de~Joannis},\ and\ \citenamefont {Holm}}]{Arnold_Holm2002b}%
  \BibitemOpen
  \bibfield  {author} {\bibinfo {author} {\bibfnamefont {A}~\bibnamefont
  {Arnold}}, \bibinfo {author} {\bibfnamefont {Jason}\ \bibnamefont
  {de~Joannis}}, \ and\ \bibinfo {author} {\bibfnamefont {Christian}\
  \bibnamefont {Holm}},\ }\bibfield  {title} {\enquote {\bibinfo {title}
  {Electrostatics in periodic slab geometries. i},}\ }\href@noop {} {\bibfield
  {journal} {\bibinfo  {journal} {J. Chem. Phys.}\ }\textbf {\bibinfo {volume}
  {117}},\ \bibinfo {pages} {2496--2502} (\bibinfo {year} {2002})}\BibitemShut
  {NoStop}%
\bibitem [{\citenamefont {Min\'{a}ry}\ \emph {et~al.}(2002)\citenamefont
  {Min\'{a}ry}, \citenamefont {Tuckerman}, \citenamefont {Pihakari},\ and\
  \citenamefont {Martyna}}]{Tuckerman_Martyna2002}%
  \BibitemOpen
  \bibfield  {author} {\bibinfo {author} {\bibfnamefont {Peter}\ \bibnamefont
  {Min\'{a}ry}}, \bibinfo {author} {\bibfnamefont {Mark~E.}\ \bibnamefont
  {Tuckerman}}, \bibinfo {author} {\bibfnamefont {Katianna~A.}\ \bibnamefont
  {Pihakari}}, \ and\ \bibinfo {author} {\bibfnamefont {Glenn~J.}\ \bibnamefont
  {Martyna}},\ }\bibfield  {title} {\enquote {\bibinfo {title} {A new
  reciprocal space based treatment of long range interactions on surfaces},}\
  }\href@noop {} {\bibfield  {journal} {\bibinfo  {journal} {J. Chem. Phys.}\
  }\textbf {\bibinfo {volume} {116}},\ \bibinfo {pages} {5351--5362} (\bibinfo
  {year} {2002})}\BibitemShut {NoStop}%
\bibitem [{\citenamefont {Pan}\ and\ \citenamefont {Hu}(2014)}]{Pan_Hu2014}%
  \BibitemOpen
  \bibfield  {author} {\bibinfo {author} {\bibfnamefont {Cong}\ \bibnamefont
  {Pan}}\ and\ \bibinfo {author} {\bibfnamefont {Zhonghan}\ \bibnamefont
  {Hu}},\ }\bibfield  {title} {\enquote {\bibinfo {title} {Rigorous error
  bounds for ewald summation of electrostatics at planar interfaces},}\
  }\href@noop {} {\bibfield  {journal} {\bibinfo  {journal} {J. Chem. Theory
  Comput.}\ }\textbf {\bibinfo {volume} {10}},\ \bibinfo {pages} {534--542}
  (\bibinfo {year} {2014})}\BibitemShut {NoStop}%
\bibitem [{\citenamefont {Bengtsson}(1999)}]{Bengtsson1999}%
  \BibitemOpen
  \bibfield  {author} {\bibinfo {author} {\bibfnamefont {Lennart}\ \bibnamefont
  {Bengtsson}},\ }\bibfield  {title} {\enquote {\bibinfo {title} {Dipole
  correction for surface supercell calculations},}\ }\href@noop {} {\bibfield
  {journal} {\bibinfo  {journal} {Phys. Rev. B}\ }\textbf {\bibinfo {volume}
  {59}},\ \bibinfo {pages} {12301--12304} (\bibinfo {year} {1999})}\BibitemShut
  {NoStop}%
\bibitem [{\citenamefont {Zhang}\ \emph {et~al.}(2020)\citenamefont {Zhang},
  \citenamefont {Sayer}, \citenamefont {Hutter},\ and\ \citenamefont
  {Sprik}}]{Zhang_Sprik2020}%
  \BibitemOpen
  \bibfield  {author} {\bibinfo {author} {\bibfnamefont {Chao}\ \bibnamefont
  {Zhang}}, \bibinfo {author} {\bibfnamefont {Thomas}\ \bibnamefont {Sayer}},
  \bibinfo {author} {\bibfnamefont {Jürg}\ \bibnamefont {Hutter}}, \ and\
  \bibinfo {author} {\bibfnamefont {Michiel}\ \bibnamefont {Sprik}},\
  }\bibfield  {title} {\enquote {\bibinfo {title} {Modelling electrochemical
  systems with finite field molecular dynamics},}\ }\href@noop {} {\bibfield
  {journal} {\bibinfo  {journal} {J. Phys.: Energy}\ }\textbf {\bibinfo
  {volume} {2}},\ \bibinfo {pages} {032005} (\bibinfo {year}
  {2020})}\BibitemShut {NoStop}%
\bibitem [{\citenamefont {D.~Wolf}\ and\ \citenamefont
  {Eggebrecht}(1999)}]{Wolf1999}%
  \BibitemOpen
  \bibfield  {author} {\bibinfo {author} {\bibfnamefont {S.R.~Phillpot}\
  \bibnamefont {D.~Wolf}, \bibfnamefont {P.~Keblinski}}\ and\ \bibinfo {author}
  {\bibfnamefont {J.}~\bibnamefont {Eggebrecht}},\ }\href@noop {} {\bibfield
  {journal} {\bibinfo  {journal} {J. Chem. Phys.}\ }\textbf {\bibinfo {volume}
  {110}},\ \bibinfo {pages} {8254} (\bibinfo {year} {1999})}\BibitemShut
  {NoStop}%
\bibitem [{\citenamefont {Chen}\ \emph {et~al.}(2004)\citenamefont {Chen},
  \citenamefont {Kaur},\ and\ \citenamefont {Weeks}}]{Chen_Weeks2004}%
  \BibitemOpen
  \bibfield  {author} {\bibinfo {author} {\bibfnamefont {Yng-gwei}\
  \bibnamefont {Chen}}, \bibinfo {author} {\bibfnamefont {Charanbir}\
  \bibnamefont {Kaur}}, \ and\ \bibinfo {author} {\bibfnamefont {John~D.}\
  \bibnamefont {Weeks}},\ }\bibfield  {title} {\enquote {\bibinfo {title}
  {Connecting systems with short and long ranged interactions: Local molecular
  field theory for ionic fluids},}\ }\href@noop {} {\bibfield  {journal}
  {\bibinfo  {journal} {J. Phy. Chem. B}\ }\textbf {\bibinfo {volume} {108}},\
  \bibinfo {pages} {19874--19884} (\bibinfo {year} {2004})}\BibitemShut
  {NoStop}%
\bibitem [{\citenamefont {Hansen}\ and\ \citenamefont
  {McDonald}(2006)}]{Hansen_McDonald2006}%
  \BibitemOpen
  \bibfield  {author} {\bibinfo {author} {\bibfnamefont {J.~P.}\ \bibnamefont
  {Hansen}}\ and\ \bibinfo {author} {\bibfnamefont {I.~R.}\ \bibnamefont
  {McDonald}},\ }\href@noop {} {\emph {\bibinfo {title} {Theory of simple
  liquids}}},\ \bibinfo {edition} {3rd}\ ed.\ (\bibinfo  {publisher} {Academic
  Press, Inc.},\ \bibinfo {address} {Amsterdam},\ \bibinfo {year}
  {2006})\BibitemShut {NoStop}%
\bibitem [{\citenamefont {Caillol}(1992)}]{Caillol1992}%
  \BibitemOpen
  \bibfield  {author} {\bibinfo {author} {\bibfnamefont {J.~M.}\ \bibnamefont
  {Caillol}},\ }\bibfield  {title} {\enquote {\bibinfo {title} {Asymptotic
  behavior of the pair‐correlation function of a polar liquid},}\ }\href@noop
  {} {\bibfield  {journal} {\bibinfo  {journal} {J. Chem. Phys.}\ }\textbf
  {\bibinfo {volume} {96}},\ \bibinfo {pages} {7039--7053} (\bibinfo {year}
  {1992})}\BibitemShut {NoStop}%
\bibitem [{\citenamefont {de~Leeuw}\ \emph {et~al.}(1980)\citenamefont
  {de~Leeuw}, \citenamefont {Perram},\ and\ \citenamefont
  {Smith}}]{DeLeeuw_Smith1980}%
  \BibitemOpen
  \bibfield  {author} {\bibinfo {author} {\bibfnamefont {S.~W.}\ \bibnamefont
  {de~Leeuw}}, \bibinfo {author} {\bibfnamefont {J.~W.}\ \bibnamefont
  {Perram}}, \ and\ \bibinfo {author} {\bibfnamefont {E.~R.}\ \bibnamefont
  {Smith}},\ }\bibfield  {title} {\enquote {\bibinfo {title} {Simulation of
  electrostatic systems in periodic boundary conditions. i. lattice sums and
  dielectric constants},}\ }\href@noop {} {\bibfield  {journal} {\bibinfo
  {journal} {Proc. R. Soc. London, Ser. A Math. Phys. Sci.}\ }\textbf {\bibinfo
  {volume} {373}},\ \bibinfo {pages} {27--56} (\bibinfo {year}
  {1980})}\BibitemShut {NoStop}%
\bibitem [{\citenamefont {Caillol}(1994)}]{Caillol1994}%
  \BibitemOpen
  \bibfield  {author} {\bibinfo {author} {\bibfnamefont {Jeanichel}\
  \bibnamefont {Caillol}},\ }\bibfield  {title} {\enquote {\bibinfo {title}
  {Comments on the numerical simulations of electrolytes in periodic boundary
  conditions},}\ }\href@noop {} {\bibfield  {journal} {\bibinfo  {journal} {J.
  Chem. Phys.}\ }\textbf {\bibinfo {volume} {101}},\ \bibinfo {pages}
  {6080--6090} (\bibinfo {year} {1994})}\BibitemShut {NoStop}%
\bibitem [{\citenamefont {Stillinger}\ and\ \citenamefont
  {Lovett}(1968)}]{Stillinger_Lovett1968}%
  \BibitemOpen
  \bibfield  {author} {\bibinfo {author} {\bibfnamefont {Frank~H.}\
  \bibnamefont {Stillinger}}\ and\ \bibinfo {author} {\bibfnamefont {Ronald}\
  \bibnamefont {Lovett}},\ }\bibfield  {title} {\enquote {\bibinfo {title}
  {General restriction on the distribution of ions in electrolytes},}\
  }\href@noop {} {\bibfield  {journal} {\bibinfo  {journal} {J. Chem. Phys.}\
  }\textbf {\bibinfo {volume} {49}},\ \bibinfo {pages} {1991--1994} (\bibinfo
  {year} {1968})}\BibitemShut {NoStop}%
\bibitem [{\citenamefont {Chandler}(1977)}]{Chandler1977}%
  \BibitemOpen
  \bibfield  {author} {\bibinfo {author} {\bibfnamefont {David}\ \bibnamefont
  {Chandler}},\ }\bibfield  {title} {\enquote {\bibinfo {title} {The dielectric
  constant and related equilibrium properties of molecular fluids: Interaction
  site cluster theory analysis},}\ }\href@noop {} {\bibfield  {journal}
  {\bibinfo  {journal} {J. Chem. Phys.}\ }\textbf {\bibinfo {volume} {67}},\
  \bibinfo {pages} {1113--1124} (\bibinfo {year} {1977})}\BibitemShut {NoStop}%
\bibitem [{\citenamefont {Rodgers}\ and\ \citenamefont
  {Weeks}(2009)}]{Rodgers_Weeks2009}%
  \BibitemOpen
  \bibfield  {author} {\bibinfo {author} {\bibfnamefont {Jocelyn~M.}\
  \bibnamefont {Rodgers}}\ and\ \bibinfo {author} {\bibfnamefont {John~D.}\
  \bibnamefont {Weeks}},\ }\bibfield  {title} {\enquote {\bibinfo {title}
  {Accurate thermodynamics for short-ranged truncations of coulomb interactions
  in site-site molecular models},}\ }\href@noop {} {\bibfield  {journal}
  {\bibinfo  {journal} {J. Chem. Phys.}\ }\textbf {\bibinfo {volume} {131}},\
  \bibinfo {pages} {244108} (\bibinfo {year} {2009})}\BibitemShut {NoStop}%
\bibitem [{\citenamefont {Hu}\ and\ \citenamefont
  {Weeks}(2010)}]{Hu_Weeks2010lmf}%
  \BibitemOpen
  \bibfield  {author} {\bibinfo {author} {\bibfnamefont {Zhonghan}\
  \bibnamefont {Hu}}\ and\ \bibinfo {author} {\bibfnamefont {John~D.}\
  \bibnamefont {Weeks}},\ }\bibfield  {title} {\enquote {\bibinfo {title}
  {Efficient solutions of self-consistent mean field equations for dewetting
  and electrostatics in nonuniform liquids},}\ }\href@noop {} {\bibfield
  {journal} {\bibinfo  {journal} {Phys. Rev. Lett.}\ }\textbf {\bibinfo
  {volume} {105}},\ \bibinfo {pages} {140602} (\bibinfo {year}
  {2010})}\BibitemShut {NoStop}%
\bibitem [{\citenamefont {Berendsen}\ \emph {et~al.}(1987)\citenamefont
  {Berendsen}, \citenamefont {Grigera},\ and\ \citenamefont
  {Straatsma}}]{Berendsen_Straatsma1987}%
  \BibitemOpen
  \bibfield  {author} {\bibinfo {author} {\bibfnamefont {H.~J.~C.}\
  \bibnamefont {Berendsen}}, \bibinfo {author} {\bibfnamefont {J.~R.}\
  \bibnamefont {Grigera}}, \ and\ \bibinfo {author} {\bibfnamefont {T.~P.}\
  \bibnamefont {Straatsma}},\ }\bibfield  {title} {\enquote {\bibinfo {title}
  {The missing term in effective pair potentials},}\ }\href@noop {} {\bibfield
  {journal} {\bibinfo  {journal} {J. Phys. Chem.}\ }\textbf {\bibinfo {volume}
  {91}},\ \bibinfo {pages} {6269--6271} (\bibinfo {year} {1987})}\BibitemShut
  {NoStop}%
\bibitem [{\citenamefont {Remsing}\ \emph {et~al.}(2016)\citenamefont
  {Remsing}, \citenamefont {Liu},\ and\ \citenamefont
  {Weeks}}]{Remsing_Weeks2016}%
  \BibitemOpen
  \bibfield  {author} {\bibinfo {author} {\bibfnamefont {R.~C.}\ \bibnamefont
  {Remsing}}, \bibinfo {author} {\bibfnamefont {S.}~\bibnamefont {Liu}}, \ and\
  \bibinfo {author} {\bibfnamefont {J.~D.}\ \bibnamefont {Weeks}},\ }\bibfield
  {title} {\enquote {\bibinfo {title} {Long-ranged contributions to solvation
  free energies from theory and short-ranged models},}\ }\href@noop {}
  {\bibfield  {journal} {\bibinfo  {journal} {Proc. Natl. Acad. Sci. USA}\
  }\textbf {\bibinfo {volume} {117}},\ \bibinfo {pages} {2819--2826} (\bibinfo
  {year} {2016})}\BibitemShut {NoStop}%
\bibitem [{\citenamefont {Gao}\ \emph {et~al.}(2020)\citenamefont {Gao},
  \citenamefont {Remsing},\ and\ \citenamefont {Weeks}}]{Gao_Weeks2020}%
  \BibitemOpen
  \bibfield  {author} {\bibinfo {author} {\bibfnamefont {Ang}\ \bibnamefont
  {Gao}}, \bibinfo {author} {\bibfnamefont {Richard~C.}\ \bibnamefont
  {Remsing}}, \ and\ \bibinfo {author} {\bibfnamefont {John~D.}\ \bibnamefont
  {Weeks}},\ }\bibfield  {title} {\enquote {\bibinfo {title} {Short solvent
  model for ion correlations and hydrophobic association},}\ }\href@noop {}
  {\bibfield  {journal} {\bibinfo  {journal} {Proc. Natl. Acad. Sci.}\ }\textbf
  {\bibinfo {volume} {117}},\ \bibinfo {pages} {1293--1302} (\bibinfo {year}
  {2020})}\BibitemShut {NoStop}%
\bibitem [{not()}]{notedetails}%
  \BibitemOpen
  \href@noop {} {}\bibinfo {note} {The complete simulation code has been
  uploaded to
  github.com/zhonghanhu1981/mdcode/zm.ionic.water.tar.gz}\BibitemShut {NoStop}%
\bibitem [{\citenamefont {Liang}\ \emph {et~al.}(2021)\citenamefont {Liang},
  \citenamefont {Xu},\ and\ \citenamefont {Zhao}}]{Liang_Xu2021}%
  \BibitemOpen
  \bibfield  {author} {\bibinfo {author} {\bibfnamefont {Jiuyang}\ \bibnamefont
  {Liang}}, \bibinfo {author} {\bibfnamefont {Zhenli}\ \bibnamefont {Xu}}, \
  and\ \bibinfo {author} {\bibfnamefont {Yue}\ \bibnamefont {Zhao}},\
  }\bibfield  {title} {\enquote {\bibinfo {title} {Random-batch list algorithm
  for short-range molecular dynamics simulations},}\ }\href@noop {} {\bibfield
  {journal} {\bibinfo  {journal} {J. Chem. Phys.}\ }\textbf {\bibinfo {volume}
  {155}},\ \bibinfo {pages} {044108} (\bibinfo {year} {2021})}\BibitemShut
  {NoStop}%
\end{thebibliography}

\end{document}